\documentclass[a4paper,11pt]{article}
\usepackage[utf8]{inputenc}

\usepackage[normalem]{ulem}
\usepackage{graphics,color}
\usepackage{csquotes}
\usepackage{amsmath,amssymb,amsthm,mathrsfs,bbm}

\newcommand{\ket}[1]{|#1\rangle  }

\title{Speedup of adiabatic multiqubit state-transfer by ultrastrong coupling of matter and radiation}

\author{M.~Stramacchia$^{1}$, A.~Ridolfo $^{2,3,5}$,  G.~Benenti $^{1,3}$,  
E.~Paladino $^{2,3,4}$, \\F.~M.~D.~Pellegrino $^{2,3}$, G.~D. Maccarrone$^{2}$  and G. Falci$^{2,3,4,*}$\\
\small$^{1}$~Center for Nonlinear and Complex Systems, Dipartimento di Scienza e Alta Tecnologia,\\
\small Universit\`a dell'Insubria, Como (Italy)\\
\small $^{2}$~Dipartimento di Fisica e Astronomia \textquote{Ettore Majorana},\\ 
\small Universit\`a di Catania (Italy)\\
\small $^{3}$~Istituto Nazionale di Fisica Nucleare, (Italy)\\
\small $^{4}$~CNR-IMM, Catania (Italy)\\
\small $^{5}$~RIKEN, Saitama, 351-0198, Tokyo (Japan)\\
\small $^*$~Correspondence: gfalci@dmfci.unict.it
}

\date{}

\begin{document}

\maketitle

\begin{abstract}
Ultrastrongly coupled quantum hardware may increase the speed of quantum state processing 
in distributed architectures, allowing to approach fault-tolerant threshold. 
We show that circuit QED architectures in the ultrastrong coupling regime, which has been recently demonstrated with superconductors, 
may show substantial speedup for a class of adiabatic protocols resilient to the main source of errors, namely the interplay of dynamical Casimir effect and cavity losses.
\end{abstract}

%


\section{Introduction}
Atom-cavity like systems in the strong coupling (SC) regime~\cite{kb:206-haroche-raimond} 
implemented by circuit-QED solid-state systems~\cite{ka:204-wallraff-superqubit} are nowadays one of the most promising platforms for quantum hardware~\cite{kr:208-schoelkopf-nature-wiring}.
Systems of superconducting artificial atoms coupled in the  SC regime to an electromagnetic resonator are at the basis of the IBM, Intel, and Google  quantum chips, the latter recently announced up to 72 qubits~\cite{kelly}. Recently entanglement has been demonstrated in systems with ten qubits~\cite{ka:217-pan-prl-tenqubitentanglement}. Circuit-QED architectures are moreover paradigm models for 
studying fundamental physics from measurement theory~\cite{ka:204-wallraff-superqubit} to quantum thermodynamics~\cite{ka:218-distefanopaternostro-prb-measurethermo} and quantum communication~\cite{ka:209-benentifalci-prl-memorych}.  
\\
We consider the simplest multiqubit architecture, two atoms coupled to a quantized mode, described by the Rabi model~\cite{kb:206-haroche-raimond,ka:216-braaksolano-jphysa-rabimodel}:
\begin{equation}\label{eq:HRabi}
H_R = \omega_c \,a^\dagger a - {1 \over 2} \,\sum_{i=1}^2 \epsilon_i \, \sigma_3^i 
+  \sum_i g^i(t) (a^\dagger\,\sigma_-^i + a \,\sigma_+^i) 
+  \sum_i g^i(t) (a\,\sigma_-^i + a^\dagger \,\sigma_+^i) 
\end{equation}
Here $\omega_c$ is the oscillator frequency and $a^\dagger$ ($a$) are the annihilation (creation) operator acting on the Hilbert space spanned by the Fock states $\{\ket{n}\}$. Pauli matrices
$\sigma^i_\alpha$ for $\alpha=3,\pm$ refer to the $i=1,2$ qubits whose splittings are 
$\epsilon_i$. The whole Hilbert space is spanned by the factorized basis $\{\ket{n\,\sigma^2\,\sigma^1}\}$, where $\ket{\sigma^i} = \{\ket{g},\ket{e}\}$ are eigenstates of $\sigma_3^i$, with eigenvalue $1$ and $-1$, respectively. Atoms-mode couplings in the dipole 
approximation~\cite{kb:206-haroche-raimond} are $g^i$, which in the SC limit $g^i$ are large enough to overcome the atomic decoherence rates ($\gamma^i$) and loss of the oscillator mode ($\kappa$). Moreover $g  \ll \omega_c$ which allows the rotating wave approximations (RWA), where the {\em counterrotating} last term of $H_R$ is neglected. The resulting $H_{RWA}$ is a Jaynes-Cummings like Hamiltonian~\cite{kb:206-haroche-raimond}, whose dynamics conserves the number of excitation $N$. Therefore it allows design of state processing with simple building block operations, namely manipulations of single excitations across the network. The simplest family is based on Rabi swaps, the cavity working as a quantum bus~\cite{ka:203-plastinafalci-prb}. Transformations depend on $g T$, where $T$ is the duration of the gate, thus increasing $g$ beyond the SC limit (typically $g \lesssim 10^{-2}\,\omega_c$) may lead to speed-up, which is important for meeting the threshold for error correction. 

\section{Multiqubit operation in the ultrastrong coupling regime}
In the last few years atom-cavity systems in the regime of ultrastrong coupling (USC), where $g \sim 0.1-1 \,\omega_c$, have been fabricated on several solid-state platforms 
(see the review~\cite{kr:217-nori-review-supercqed}). In the simplest instance they are described by the full Rabi Hamiltonian Eq.(\ref{eq:HRabi}).
The counterrotating terms spoil conservation of $N$. A remarkable feature is that large values of $g^1$ together with a modulation in time determine the production of photon pairs, a phenomenon known as the dynamical Casimir effect (DCE). As a consequence the USC quantum bus is sensitive to multiphoton generation, and DCE deteriorates the fidelity of quantum operations~\cite{ka:214-benenti-pra-casmirrabi} even in absence of decoherence. This poses a fundamental limit for standard protocols based on Rabi oscillations.

\begin{figure}[t!]
\centering
\resizebox{0.9\textwidth}{0.18\textheight}{\includegraphics{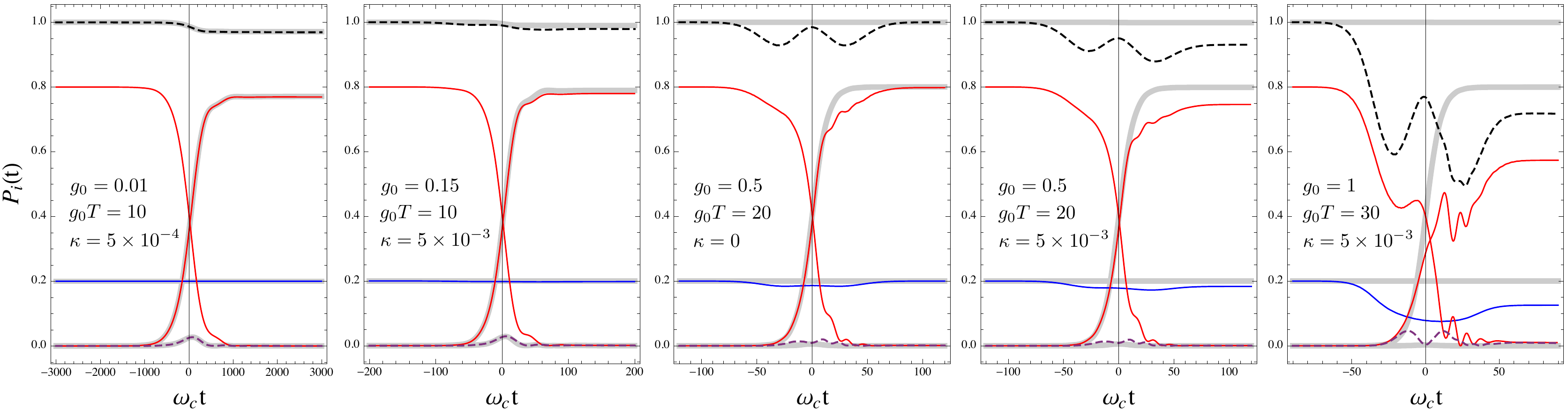}}
\caption{\small Population transfer by STIRAP in RWA and USC, at the doubly resonant point $\omega_c = \epsilon_i$, for  
initial state $\ket{\psi_0} = \sqrt{0.2} \ket{0gg} + i \sqrt{0.8} \ket{0ge}$ and $\tau = 0.7\, T$.   
Colors: blue ($P_{\ket{0gg}}$), red ($P_{\ket{0ge}}$ and $P_{\ket{0eg}}$), magenta dashed 
($P_{\ket{1gg}}$), black dashed (sum of the above populations, i.e. the STIRAP subspace), thick gray (RWA). Parameters are in units of $\omega_c$. Counterrotating terms do not modify the RWA behavior up to $g_0\sim 0.15\, \omega_c$, which is a nearly optimal value for minimizing $T$; notice that a  larger $\kappa=0.005$ is used. At stronger coupling adiabaticity requires a less favorable scaling of $T$, and leakage from the STIRAP subspace occurs; this leakage is reversible for $\kappa=0$ (due to the symmetries of the Hamiltonian) but becomes detrimental for finite 
$\kappa$.
\label{fig:stirap}}
\end{figure}

To overcome this problem we propose a communication channel implemented by an adiabatic protocol similar to STIRAP~\cite{kr:217-vitanovbergmann-rmp}. Implementation of STIRAP in individual 
artificial atoms has been proposed~\cite{ka:209-siebrafalci-prb,ka:216-distefano-pra-twoplusone} in the last few years.
It has been recently demonstrated in the Transmon design~\cite{ka:216-kumarparaoanu-natcomm-stirap,ka:216-xuhanzhao-natcomm-ladderstirap}, and extended to achieve hybrid control of 
qutrits~\cite{ka:216-vepsalainen-photonics-squtrit}. We first consider $H_{RWA}$~\cite{ka:217-wangchesi-njp-optstirap,ka:217-falci-fortphys-fqmt}: the system is prepared in $\ket{0ge}$ i.e. qubit $1$ is in the excited state, and we switch on $g^{(2)}(t) = g \, \mathrm{e}^{- [(t+\tau)/T]^2}$ {\em before} 
$g^{(1)}(t) = g \, \mathrm{e}^{- [(t-\tau)/T]^2}$. The protocol will end in $\ket{0eg}$, provided $g T \gg 1$ to guarantee adiabaticity of the dynamics (in practice, $gT>5$ is sufficient). If this is 
the case the intermediate state  $\ket{1gg}$ is never populated during the whole protocol, thus the cavity acts as a {\em virtual quantum bus}. 
The protocol leaves unaffected the ground state $\ket{0gg}$, 
thus $H_{RWA}$ would implement ideal state transfer $\ket{0g} (\alpha \ket{g} + \beta \ket{e}) \to  \ket{0} (\alpha \ket{g} + \beta \ket{e}) \ket{g}$ and other two qubit operations can be implemented with $\sim 100 \%$ efficiency.
One may wonder if in the USC regime STIRAP brings advantages over protocols based on Rabi swaps,
in view of the fact that the cavity is almost never populated which should reduce the impact 
of DCE and photon losses.

\begin{figure}[t!]
\centering
\resizebox{!}{0.15\textheight}{\includegraphics{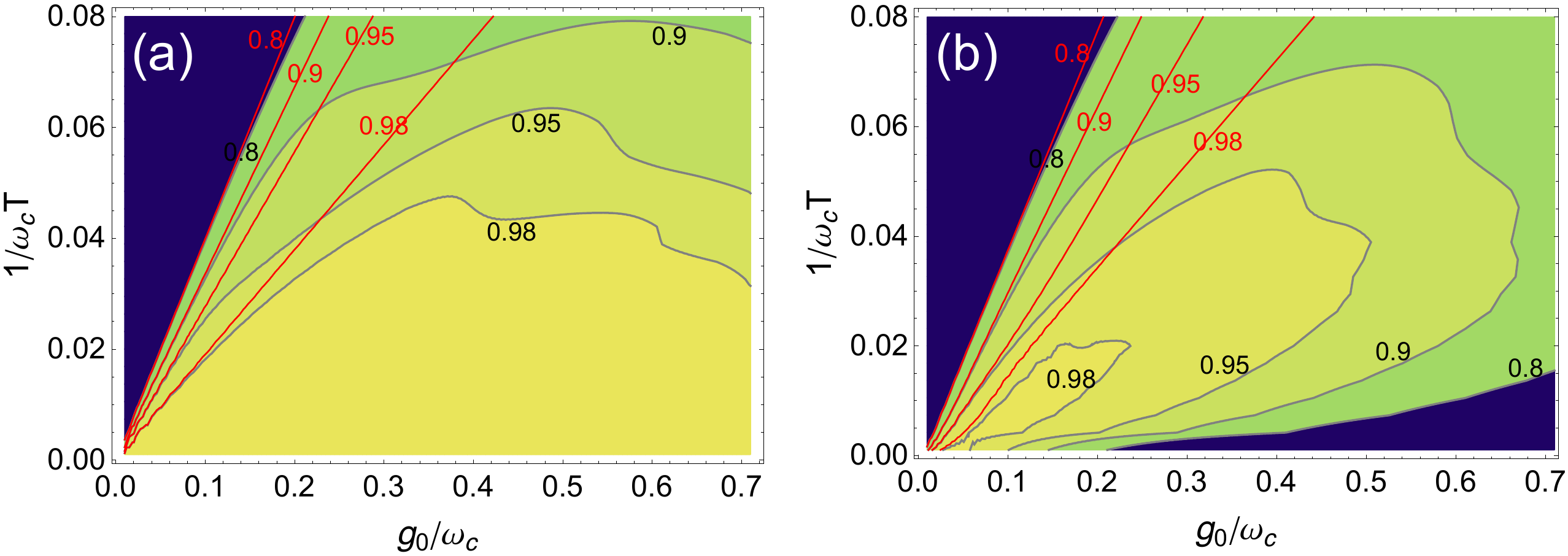}}
\resizebox{!}{0.17\textheight}{\includegraphics{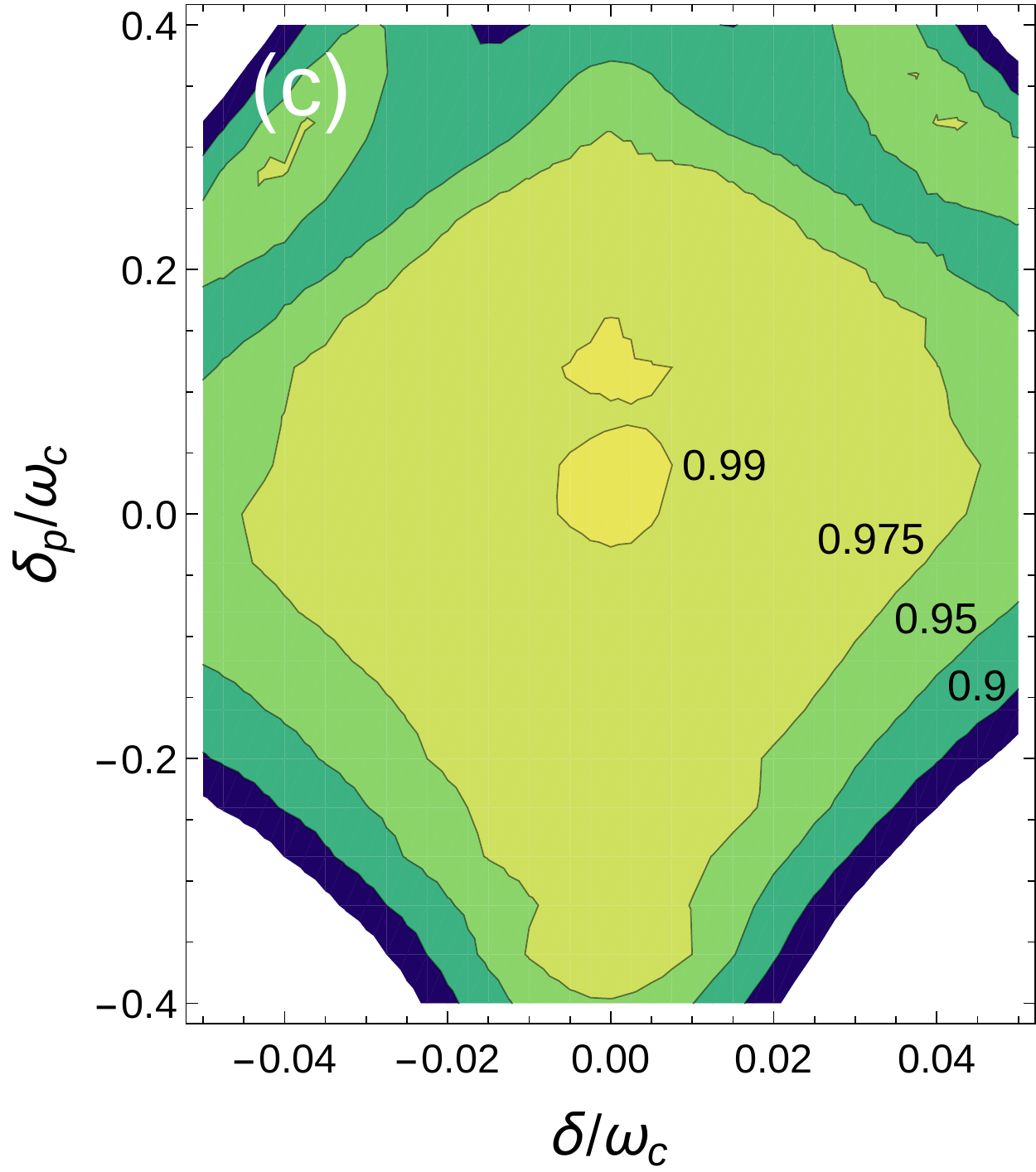}}
\caption{\small (a) Efficiency of population transfer $P_{\ket{0eg}}(t_f)$ in the $N=1$ manifold for Rabi model and for the RWA (straight lines) for 
$\kappa=0$. (b) The same for $\kappa=0.005$, showing a region around $g/\omega_c=0.15$ where one may take advantage of the ultrastrong coupling, while at larger $g$ combination of DCE and decay spoil the efficiency. 
(c) Sensitivity of population transfer to detunings, showing that the protocol is robust against fluctuations in a wide region about the resonance point $\delta=\delta_p=0$, and that positive 
single photon detuning $\delta_p>0$ may further increase the efficiency. 
\label{fig:efficiency}}
\end{figure}

We first study population transfer from the initial state 
$\ket{0g} (\alpha \ket{g} + \beta \ket{e})$ at the doubly resonant point
$\omega_c = \epsilon_i$. In RWA the dynamics depends only on $gT$ therefore the final population $P_i(t_f|g_0T)$, is constant on each line $T = A/g_0$ 
(see Fig.~\ref{fig:efficiency}a). Larger $A$ guarantees adiabaticity, the 
efficiency of population transfers being $\sim 1$ for $A>5$.
In the USC regime, with cavity losses modeled by the replacement $\omega_c \to \omega_c - i \kappa$,
results in Fig.~\ref{fig:stirap} shows that: 
(1) the final population of $\ket{0gg}$ is unaffected by USC even if during the protocol it is 
(blue curve); (2) population from $\ket{0ge}$ is transferred in 
$\ket{0eg}$ (red curves); (3) the main imperfection appearing 
for $g/\omega_c \lesssim 0.2$ is population of $\ket{1gg}$ at intermediate times 
(magenta curve), whereas for larger $g \gtrsim 0.5$ 
leakage from the $N=0,1$ subspaces (related to DCE) is the dominant stray effect. 

For increasing $g$, the protocol becomes faster. Notice in particular that passing from the SC regime with $g=0.01$ regime to USC with $g=0.5$, the typical time scale for the protocol reduces 
from $T\approx 1000$ to  $T\approx 40$ (in units of $\omega_c$). However, 
while the needed $T$ scales as $1/g$ up to $g\sim 0.2$, further increase 
of $g$ brings no advantage, since leakage to the $N=2,3$ subspaces becomes relevant. 
Although the STIRAP protocol would be by itself resilient to leakage due to symmetries in the
Hamiltonian, this is no longer true in the presence of cavity losses, the main source 
of decoherence in present USC systems. Then leakage becomes irreversible and produces a substantial loss in the fidelity of population transfer. 

In realistic cases $\kappa$ increases with $g$ but the protocol requires smaller time. Therefore
there will be a tradeoff between speedup and errors due to the interplay of DCE and losses. 
To address this point we compare state transfer in the USC regime and in RWA, 
characterizing the impact of DCE plus losses as a function of $g$ and $T$.
Fig.~\ref{fig:efficiency}a shows the efficiency of population transfer in the $N=1$ subspace 
for $\kappa=0$. For $H_{RWA}$ it displays the dependence on $gT$ only. This behavior approximately holds in the USC regime, the DCE reducing the efficiency for $g > 0.3$. The favorable region shrinks due to the interplay with cavity losses (Fig.~\ref{fig:efficiency}b, where $\kappa=0.005$), but still in a region around $g \sim  0.15$ one may take advantage from USC. In this region we then study robustness of population transfer against parametric fluctuations, which is one of the main assets of STIRAP. Fig.~\ref{fig:efficiency}c shows that the protocol keeps a very large efficiency  
in region around the doubly resonant point where the single-photon detuning 
$\delta_p = \epsilon_1-\omega_c$ and the two-photon detuning $\delta= \epsilon_1-\epsilon_2$ are nonvanishing. Notice that since the energy scale in the figure is $\omega_c$ this stability region is very large. Therefore the protocol is very robust, a feature essentially due to the USC between artificial atom and cavity. Another important property is that using 
non-vanishing single-photon detuning $\delta_p > 0$ the efficiency is further increased. Both features have important consequences that we briefly discuss in the conclusions.

\section{Conclusion}
We have shown that a protocol based on STIRAP allows ultrafast and reliable state transfer in the USC regime, up to relatively large values of the qubits-field coupling strength $g \sim 0.2$, 
in spite of the joint presence of DCE, inducing leakage from the computational subspaces, and of 
large cavity losses determining Markovian decoherence. On the contrary protocols based on Rabi swaps are very sensitive to DCE~\cite{ka:214-benenti-pra-casmirrabi}, noise further degrading performances. In addition the STIRAP-based protocol is much more robust, a detailed study of population transfer has showing that very large efficiencies $> 99\%$ may be reached in a wide region of the space of the parameters. This property is essentially due to the large coupling strength achieved in the USC regime, and it has the important consequence of rendering our protocol resilient also to low-frequency noise~\cite{kr:214-paladino-rmp,ka:213-falci-prb-stirapcpb}. Since STIRAP-based protocols may also be used to generate entanglement, this resilience may provide an alternative strategy to fight decoherence in two-qubit gates~\cite{ka:211-paladino-njp-univ2qubit}.  
Therefore, it would be interesting to investigate other STIRAP-inspired protocols for multiqubit
entanglement generation in the USC regime, and compare with other existing proposals~\cite{ka:214-felicetti-prl-dceentanglement}. Another important feature is that we have shown that efficiency may be further increased by using a non-vanishing single-photon atom-cavity detuning. This feature is reminiscent to what happens in a protocol called hyperstirap~\cite{kr:217-vitanovbergmann-rmp}, where efficiency can be further increased by a proper phase modulation~\cite{ka:216-distefano-pra-twoplusone} of the control. Therefore  it can be foreseen that optimal control techniques \cite{simone,DCEoptimal} could further improve the fidelity of such operations, also exploiting coupling schemes based on modulation of detunings~\cite{ka:215-distefano-prb-cstirap,ka:217-falci-fortphys-fqmt}.



\begin{thebibliography}{-------}

\bibitem{kb:206-haroche-raimond}
Haroche, S.; Raimond, J.M.
\newblock {\em Exploring the Quantum}; Oxford University press,  2006.

\bibitem{ka:204-wallraff-superqubit}
Wallraff, A.; Schuster, D.I.; Blais, A.; Frunzio, L.; Huang, R.S.; Majer, J.;
  Kumar, S.; Girvin, S.; Schoelkopf, R.
\newblock Strong coupling of a single photon to a superconducting qubit using
  circuit quantum electrodynamics.
\newblock {\em Nature} {\bf 2004}, {\em 421},~162--167.


\bibitem{kr:208-schoelkopf-nature-wiring}
Schoelkopf, R.J.; Girvin, S.M.
\newblock Wiring up quantum systems.
\newblock {\em Nature} {\bf 2008}, {\em 451},~664.

\bibitem{kelly}
Kelly, J.
\newblock Engineering superconducting qubit arrays for quantum supremacy.
\newblock {\em American Physical Society March Meeting, Los Angeles} {\bf  2018}.

\bibitem{ka:217-pan-prl-tenqubitentanglement}
Song, C.; Xu, K.; Liu, W.; Yang, C.p.; Zheng, S.B.; Deng, H.; Xie, Q.; Huang,
  K.; Guo, Q.; Zhang, L.; Zhang, P.; Xu, D.; Zheng, D.; Zhu, X.; Wang, H.;
  Chen, Y.A.; Lu, C.Y.; Han, S.; Pan, J.W.
\newblock 10-Qubit Entanglement and Parallel Logic Operations with a
  Superconducting Circuit.
\newblock {\em Phys. Rev. Lett.} {\bf 2017}, {\em 119},~180511.

\bibitem{ka:218-distefanopaternostro-prb-measurethermo}
Di~Stefano, P.G.; Alonso, J.J.; Lutz, E.; Falci, G.; Paternostro, M.
\newblock Nonequilibrium thermodynamics of continuously measured quantum
  systems: A circuit QED implementation.
\newblock {\em Phys. Rev. B} {\bf 2018}, {\em 98},~144514.

\bibitem{ka:209-benentifalci-prl-memorych}
Benenti, G.; D'Arrigo, A.; Falci, G.
\newblock Enhancement of Transmission Rates in Quantum Memory Channels with
  Damping.
\newblock {\em Phys. Rev. Lett.} {\bf 2009}, {\em 103},~020502.

\bibitem{ka:216-braaksolano-jphysa-rabimodel}
Braak, D.; Chen, Q.H.; Batchelor, M.T.; Solano, E.
\newblock Semi-classical and quantum Rabi models: in celebration of 80 years.
\newblock {\em Journal of Physics A: Mathematical and Theoretical} {\bf 2016}, {\em 49},~300301.

\bibitem{ka:203-plastinafalci-prb}
Plastina, F.; Falci, G.
\newblock Communicating Josephson qubits.
\newblock {\em Phys. Rev. B} {\bf 2003}, {\em 67}.

\bibitem{kr:217-nori-review-supercqed}
Gu, X.; Frisk~Kockum, A.; Miranowicz, A.; Liu, Y.X.; Nori, F.
\newblock Microwave photonics with superconducting quantum circuits.
\newblock {\em Physics Reports} {\bf 2017}, {\em 718–719},~1--102.
\newblock arXiv:1707.02046.

\bibitem{ka:214-benenti-pra-casmirrabi}
Benenti, G.; D’Arrigo, A.and~Siccardi, S.; Strini, G.
\newblock {\em Phys. Rev. A} {\bf 2014}, {\em 90},~052313.

\bibitem{kr:217-vitanovbergmann-rmp}
Vitanov, N.V.; Rangelov, A.A.; Shore, B.W.; Bergmann, K.
\newblock Stimulated Raman adiabatic passage in physics, chemistry, and beyond.
\newblock {\em Rev. Mod. Phys.} {\bf 2017}, {\em 89},~015006.

\bibitem{ka:209-siebrafalci-prb}
Siewert, J.; Brandes, T.; Falci, G.
\newblock Advanced control with a Cooper-pair box: Stimulated Raman adiabatic
  passage and Fock-state generation in a nanomechanical resonator.
\newblock {\em Phys. Rev. B} {\bf 2009}, {\em 79},~024504.

 \bibitem{ka:216-distefano-pra-twoplusone}
Di~Stefano, P.G.; Paladino, E.; Pope, T.J.; Falci, G.
\newblock Coherent manipulation of noise-protected superconducting artificial
  atoms in the Lambda scheme.
\newblock {\em Phys. Rev. A} {\bf 2016}, {\em 93},~051801.

\bibitem{ka:216-kumarparaoanu-natcomm-stirap}
Kumar, K.; Veps\"al\"ainen, A.; Danilin, S.; Paraoanu, G.
\newblock Stimulated Raman adiabatic passage in a three-level superconducting
  circuit.
\newblock {\em Nat. Comm.} {\bf 2016}, {\em 7},~10628.

\bibitem{ka:216-xuhanzhao-natcomm-ladderstirap}
Xu, H.; C.~Song, W.Y.; Liu, G.M.; Xue, F.F.; Su, H.; Deng, Y.; Tian, D.N.;
  Zheng, S.; Han, Y.P.; Zhong, H.; Wang, Y.x.L.; Zhao, S.P.
\newblock Coherent population transfer between uncoupled or weakly coupled
  states in ladder-type superconducting qutrits.
\newblock {\em Nature Communications} {\bf 2016}, {\em 7},~11018.

 \bibitem{ka:216-vepsalainen-photonics-squtrit}
Veps\"al\"ainen, A.; Danilin, S.; Paladino, E.; Falci, G.; Paraoanu, G.S.
\newblock Quantum Control in Qutrit Systems Using Hybrid Rabi-STIRAP Pulses.
\newblock {\em Photonics} {\bf 2016}, {\em 3},~62.

  \bibitem{ka:217-wangchesi-njp-optstirap}
Wang, Y.D.; Zhang, R.; Yan, X.B.; Chesi, S.
\newblock Optimization of STIRAP-based state transfer under dissipation.
\newblock {\em New Journal of Physics} {\bf 2017}, {\em 19},~093016.

  \bibitem{ka:217-falci-fortphys-fqmt}
Falci, G.; Di~Stefano, P.; Ridolfo, A.; D'Arrigo, A.; Paraoanu, G.; Paladino,
  E.
\newblock Advances in quantum control of three-level superconducting circuit
  architectures.
\newblock {\em Fort. Phys.} {\bf 2017}, {\em 65},~1600077.

  \bibitem{kr:214-paladino-rmp}
Paladino, E.; Galperin, Y.; Falci, G.; Altshuler, B.
\newblock 1/f noise: implications for solid-state quantum information.
\newblock {\em Rev. Mod. Phys.} {\bf 2014}, {\em 86},~361--418.

  \bibitem{ka:213-falci-prb-stirapcpb}
Falci, G.; La~Cognata, A.; Berritta, M.; D'Arrigo, A.; Paladino, E.; Spagnolo,
  B.
\newblock Design of a Lambda system for population transfer in superconducting
  nanocircuits.
\newblock {\em Phys. Rev. B} {\bf 2013}, {\em 87},~214515--1,214515--13.

  \bibitem{ka:211-paladino-njp-univ2qubit}
Paladino, E.; D'Arrigo, A.; Mastellone, A.; Falci, G.
\newblock Decoherence times of universal two-qubit gates in the presence of
  broad-band noise.
\newblock {\em New Jour. Phys.} {\bf 2011}, {\em 13}, ), 093037.

  \bibitem{ka:214-felicetti-prl-dceentanglement}
Felicetti, S.; Sanz, M.; Lamata, L.; Romero, G.; Johansson, G.; Delsing, P.;
  Solano, E.
\newblock Dynamical Casimir Effect Entangles Artificial Atoms.
\newblock {\em Phys. Rev. Lett.} {\bf 2014}, {\em 113},~093602.

\bibitem{simone}
Rach, N.; M\"uller, M.M.; Calarco, T.; Montangero, S.
\newblock Dressing the chopped-random-basis optimization: A bandwidth-limited
  access to the trap-free landscape.
\newblock {\em Phys. Rev. A} {\bf 2015}, {\em 062343},~92.

  \bibitem{DCEoptimal}
Hoeb, F.; Angaroni, F.; Zoller, J.; Calarco, T.; Strini, G.; Montangero, S.;
  Benenti, G.
\newblock Amplification of the parametric dynamical Casimir effect via optimal
  control.
\newblock {\em Phys. Rev. A} {\bf 2017}, {\em 033851},~96.

  \bibitem{ka:215-distefano-prb-cstirap}
Di~Stefano, P.G.; Paladino, E.; D'Arrigo, A.; Falci, G.
\newblock Population transfer in a Lambda system induced by detunings.
\newblock {\em Phys. Rev. B} {\bf 2015}, {\em 91},~224506.

\end{thebibliography}
\end{document}